\documentclass[oldversion, printer]{aa}

\usepackage{graphicx}
\usepackage{txfonts}
\usepackage{natbib}
\usepackage{color}
\bibpunct{(}{)}{;}{a}{}{,}

\newcommand{\up}{u}
\newcommand{\low}{\ell}

\newcommand{\ie}{i.e.,\ }
\newcommand{\eg}{e.g.,\ }

\definecolor{gray}{rgb}{.6,.6,.6}
\definecolor{green}{rgb}{0,.6,0}
\definecolor{red}{rgb}{0,0,.6}
\newcommand*{\cancel}[1]{\textcolor{gray}{}}

\begin{document}

\title{
  Internal resonance in non-linear disk oscillations and 
  the amplitude evolution of neutron star kilohertz QPOs
  }

\author{
  Ji\v{r}\'{\i} Hor\'ak\inst{1}
  \and
  Marek.\,A.~Abramowicz\inst{2,3}
  \and
  W{\l}odek Klu\'zniak\inst{4,3}
  \and
  Paola Rebusco\inst{5}
  \and
  Gabriel T\"{o}r\"{o}k\inst{6}
  }
  
\offprints{horak@astro.cas.cz}

\institute{
  Astronomical Institute of the Academy of Sciences, Bo\v{c}n\'{\i}~II 1401/1a, 141-31 Praha~4, CZ
  \and
  Department of Physics, G\"oteborg University, S-412 96 G\"oteborg, SE
  \and
  Copernicus Astronomical Centre PAN, Bartycka 18, 00-716 Warsaw, PL
  \and
  Johannes Kepler Institute of Astronomy, Zielona G\'ora University, ul. Lubuska 2, 65-265 Zielona G\'ora, PL 
  \and 
  MIT Kavli Institute for Astrophysics and Space Research, 77 Massachusetts Avenue, Cambridge, MA 02139, USA
  \and
  Institute of Physics, Faculty of Philosophy and Science, Silesian University in Opava, 
  Bezru\v{c}ovo n\'{a}m.  13,746-01 Opava, CZ
  }

\authorrunning{J.~Hor\'ak et al.}

\titlerunning{Internal resonance in neutron star kilohertz QPOs}

\date{Received .. ............. 2008; accepted .. ............. 2008}

\abstract{
We explore some properties of twin kilohertz quasiperiodic oscillations (QPOs) in a simple toy-model consisting of two  oscillation modes coupled by a general nonlinear force. We examine resonant effects by slowly varying the values of the tunable, and nearly commensurable, eigenfrequencies. The behavior of the actual oscillation frequencies and amplitudes during a slow transition through the 3:2 resonance is examined in detail and it is shown that both are significantly affected by the nonlinearities in the governing equations. In particular, the amplitudes of oscillations reflect a resonant exchange of energy between the modes, as a result the initially weaker mode may become dominant after the transition. We note that a qualitatively similar behavior has been recently reported in several neutron star sources by T\"{o}r\"{o}k (2008), who found that the difference of amplitudes in neutron star twin peak QPOs changes sign as the observed frequency ratio of the QPOs passes through the value 3:2.}


\maketitle


\section{Introduction}
\label{intro}

There is little doubt that kHz QPOs observed in low mass X-ray binaries originate in the inner region of accretion disks of the neutron star or black hole sources. The very fact that these quasi-periodic modulations are observed  in the X-ray flux, i.e., in the most luminous part of the source spectrum---and at the same time at high photon energies---implies an origin in the inner part of the accretion flow, where most of the gravitational potential energy of the accreting matter is released. Time-of-flight arguments also imply that a kHz modulation of unbeamed radiation must originate in a region not larger than $\sim100\,$km, with an even sharper bound when the high degree of coherence of the observed modulation is taken into account. 

In the present paper we focus on kHz QPOs that show up in pairs, the so called twin peak QPOs \citep[see][for a review]{Klis2000}, and we assume that they correspond to oscillations of an accretion-disk structure. Reproducible features of these high-frequency QPOs imply that they reflect some fundamental properties of the systems, such as their mass and spin. 

Many models have been proposed, some involving orbital motions \citep{Kluzniak+1990, Stella+Vietri1999}, others based on oscillations of the accretion disk \citep{Wagoner1999, Wagoner+2001, Kato2001, Kato2004, Kato2005a, Kato2005b, Kato2008, Petri2005, Petri2006} and of tori \citep{Rezzolla+2003a, Rezzolla+2003b, Bursa+2004, Montero+2004, Montero+2007, Blaes+2006, Blaes+2007, Schnittman+Rezzola2006, Zanotti+2005, Horak2008}. Most of these  models involve two or more oscillatory motions, which may interact in a non-linear manner.

The appearance of rational ratios (mainly $\sim$3:2) between the centroid of twin peak QPOs points at a nonlinear coupling between different modes of oscillations \citep{Kluzniak+Abramowicz2001a}. The importance of a resonant process was also pointed out by \citet{Psaltis+Norman2000, Kluzniak+Abramowicz2001b, Titarchuk2002, Kluzniak+2004} and investigated in more detail, e.g., in \citet{Abramowicz+2003b, Rebusco2004, Horak+2004, Lee+2004, Horak+Karas2006}.

One example of an investigation of resonances between specific modes in an accretion disk can be found in the work of \citet{Kato2005b}, who proposed a model in which the two kHz QPOs are due to a nonlinear resonant interaction between the oscillations of a thin accretion disk and an oscillatory perturbation induced by the non-axisymmetric rotating neutron star. Both kHz oscillations are localized close to a resonant radius, where the interaction with the perturbation is most prominent. The spatial proximity of the two oscillations may facilitate an internal resonance between them. As pointed out recently by \citet{Mao+2008}, a resonance may occur between two modes with non-overlapping wavefunctions, when it is mediated by a traveling wave. In other words, the two oscillators in resonance need not be located in the same annulus of the accretion disk. More recently \citet{Kato2008} relaxed the local approximation and analyzed the nonlinear resonant excitation of inertial-acoustic disk oscillations in deformed disks (a warped disk or a one-armed pattern symmetric with respect to the equator).

Other QPO models involve vortices in the disk or magnetic interactions. For example \citet{Tagger+Varniere2006} suggest the formation of two or three Rossby-wave vortices in microquasar disks, while \citet{Li+Narayan2004} suggested a model in which the two QPOs are identified with two Rayleigh-Taylor or Kelvin-Helmholtz unstable modes developed at the interface between the disk and the magnetosphere. Similarly, in the Alfv\'{e}n wave oscillations model \citep{Zhang2004, Zhang+2007} the two QPOs correspond to the Keplerian orbital frequency and the frequency of the Alfv\'{e}n waves excited at the same radius. It is less clear whether the resonant interaction described in this paper would apply to those models as well.

In this paper we consider some general properties of non-linear resonance that may be applicable to many systems. We specify neither the physical nature of the oscillators, nor the couplings necessary for a resonance to occur. In the context of kHz QPOs, the oscillators could be taken to be eigenmodes of the accretion disk or of an associated structure, perhaps ones already discussed in the extensive theoretical literature cited above. We do not address the crucial question of the excitation and damping of the modes in the turbulent disk, particularly when MRI-induced turbulence is present (in the framework of two coupled oscillators this issue was adressed already by \citet{Vio+2006}). Eventually, one would hope to identify such resonances in numerical models of accretion flow. It is only recently that inertial modes have started showing up in numerical simulations  \citep{Katoy2004, Brandenburg2005, Arras+2006, Reynolds+2008, Machida+2008}. It may be premature to look for a numerical validation of the idea of resonance in accretion disks, at least until two distinct frequencies in a 3:2 ratio are identified with some regularity in the simulations. Nevertheless, numerical work may already be offering new insights.
 
In what follows we review some general properties of internal resonance and investigate the amplitude evolution of the oscillations by modeling a slow passage of the frequencies through the 3:2 resonance. The results presented in this paper are more relevant to neutron-star QPOs where the frequencies wander over hundreds of hertz on the time-scale of several hours, than to black hole QPOs, whose frequencies are more stable.

\section{Mathematics of the internal resonance}
\label{sec:dyn}

We model the QPOs using two coupled nonlinear oscillators governed by the equations
\begin{eqnarray}
  \ddot{x}_\low + \omega_\low^2 x_\low &=& 
  \omega_\low^2 f_\low(x_\low,x_\up,\dot{x}_\low,\dot{x}_\up),
  \label{eq:govl}\\
  \ddot{x}_\up + \omega_\up^2 x_\up &=&
  \omega_\up^2 f_\up(x_\low,x_\up,\dot{x}_\low,\dot{x}_\up).
  \label{eq:govu}
\end{eqnarray}
The oscillations underlying the upper and lower QPOs are identified with the dimensionless solutions $x_{\up}(t)$ and $x_\low(t)$. The coupling functions $f_\low$ and $f_\up$ are treated as a perturbation, and are further assumed to contain only  nonlinear terms and to be invariant under time inversion (\eg their Taylor expansions start with the second order and do not contain odd powers of the time  derivatives $\dot{x}_{\low}$, $\dot{x}_\up$). The coefficients $\omega_\low>0$, and $\omega_\up>0$ are the eigenfrequencies of the two oscillators.

In order to study the most relevant resonant effects, we concentrate on the case when $\omega_{\up}/\omega_\low$ is close to 3:2. We chose this resonance as it is most prominent in the QPO data \citep{Abramowicz+2003a, Torok+2008}. The degree of closeness to the rational ratio is expressed by the detuning parameter $\sigma\equiv 2\omega_\up - 3\omega_\low$.

Our analysis is applicable when the amplitudes of oscillations and the detuning parameter are small and are such that $a_\low \sim a_\up \sim\epsilon\ll 1$ and $\sigma\sim\epsilon^2$ respectively. In that case, perturbation techniques can be used to find approximate solutions of the equations (\ref{eq:govl}) and (\ref{eq:govu}). Using the method of multiple scales \citep[e.g.\ ][]{Nayfeh+Mook1979} we find the lowest order real displacements in the form:
\begin{eqnarray}
  x_\low(t) = \Re\left[A_\low(t)e^{i\omega_\low t}\right],
  \quad
  x_\up(t) = \Re\left[A_\up(t)e^{i\omega_\up t}\right].
  \label{eq:solu}
\end{eqnarray}
The higher-order terms add only higher harmonics to the quasiperiodic signal described by these solutions. The main difference with respect to the solutions of a linear system are that the complex amplitudes 
\begin{equation}
  A_\low(t)\equiv\frac{1}{2}a_\low(t)e^{i\phi_\low(t)}, 
  \quad
  A_\up(t)\equiv\frac{1}{2}a_\up(t)e^{i\phi_\up(t)}
\end{equation}
depend on time. 
We will refer to the actual frequencies of the solution, $\omega_\low^\star$ and $\omega_\up^\star$, as the ``observed frequencies.'' They are shifted with respect to the eigenfrequencies by the corrections $\Delta\omega_\low=\dot{\phi}_\low$ and $\Delta\omega_\up=\dot{\phi}_\up$.

\begin{figure}
  \begin{center}
    \includegraphics[width=0.5\textwidth]{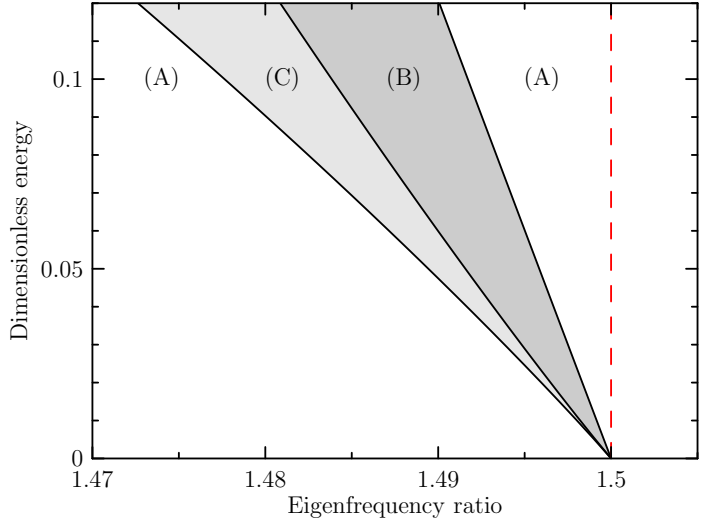}
  \end{center}
  \caption{Regions where a resonance is present (resonance `tongues')
  in the plane of the eigenfrequency ratio and the total energy of oscillations.
  Each point in the plane corresponds to one single $(\gamma,\xi)$-disk 
  and different regions to different topologies of these disks
 (see Fig.~\ref{fig:gxplanes}). Regions (A), (B) and
  (C) correspond to zero, one, and two fixed points in the $(\gamma,\xi)$-disk. 
  The parameters of the system are given in the text.}
  \label{fig:tongue}
\end{figure}

We assume energy conservation, so the two real amplitudes may be parameterized using the fractional energy $\xi(t)$:
\begin{equation}
  a_\low^2=\xi E, 
  \quad
  a_\up^2=(1-\xi)\frac{E}{\nu}, 
  \quad
  0\leq\xi\leq1,
  \label{eq:xi}
\end{equation}
where $E\equiv a_\low^2+\nu a_\up^2$ is the total dimensionless energy and $\nu$ is a constant that depends on the properties of the system \citep{Horak+Karas2006}. Similarly, we introduce the phase function $\gamma(t)$ as
\begin{equation}
  \gamma\equiv2\phi_\up-3\phi_\low+\sigma t.
\end{equation}
The difference $2\omega_\up^\star - 3\omega_\low^\star$ expressed using the frequency
corrections is
\begin{equation}
  2\omega_\up^\star - 3\omega_\low^\star = \sigma - 3\dot{\phi}_\low +
  2\dot{\phi}_\up = \dot{\gamma}.
\end{equation}
Hence, the observed frequencies are commensurable whenever $\dot{\gamma}=0$, even if the eigenfrequencies are not exactly so (the deviation is proportional to $\epsilon^{~2}$).  

The dynamics of the system is then governed by the two first-order ordinary differential equations
\begin{eqnarray}
  \dot{\xi} \!\!&=&\!\! \frac{1}{8}\beta\omega_\up(1-\xi)(\xi
  E)^{3/2}\sin\gamma,
  \label{eq:govx}\\
  \dot{\gamma} \!\!&=&\!\! \sigma + \frac{1}{4}\,\omega_\up E\,
  \Big[\mu_\low\,\xi +
   \frac{\mu_\up}{\nu}(1-\xi)
   +\frac{1}{4}\,
  \beta(3-5\xi)(\xi E)^{1/2}\cos\gamma\Big],
  \label{eq:govg}
\end{eqnarray}
where $\beta$, $\mu_\up$ and $\mu_\low$ are as yet unspecified dimensionless constants characterizing a given system. 

The values of the system parameters follow directly from the coefficients of the Taylor expansions of the functions $f_{\low}$ and $f_\up$ and depend on the particular physical model. For example, \citet{Rebusco2004} studied small nonlinear epicyclic oscillations in nearly geodesic orbits of test particles in Schwarzschild spacetime using the method of multiple scales. Later, \citet{Horak+Karas2006} generalized her approach to motion in an arbitrary axisymmetric gravitational field. In both works the radial and vertical oscillations of the particles are governed by equations of the form of (\ref{eq:govl}) and (\ref{eq:govu}), \eg equations (14) and (15) in \citet{Horak+Karas2006}. Moreover, the latter work contains explicit formulae for the constants $\beta$, $\mu_\up$ and $\mu_\low$ in terms of derivatives of the effective potential [see the relations after their equation (68)]. In addition, a general form of the equations (\ref{eq:govl}) and (\ref{eq:govu}) is suitable for studying resonance phenomena in different physical conditions. \citet{Horak2008} examined nonlinear interaction between two epicyclic modes of a slender accretion torus; sequation (6) in his paper represents a particular example of the general equations discussed here.

Both, $\xi(t)$ and $\gamma(t)$, are slowly varying functions of time, they describe a slow modulation of amplitudes and phases during the oscillations. The amplitude modulation occurs on the characteristic time scale $t_\mathrm{mod}\equiv\epsilon^{-3}t_\mathrm{osc}$, with $t_\mathrm{osc}\equiv\omega_\up^{-1}$ being the characteristic timescale of the oscillations. The two amplitudes are anticorrelated and the variations of $\xi$ correspond to an exchange of energy between the modes that keeps the total energy of the oscillations, $E$, constant. Since the oscillations are nonlinear, the observed frequencies are periodically modulated reflecting the modulation of the amplitudes. In the context of QPOs this process has already been discussed by \citet{Horak+2004} in connection to the correlations between the QPO frequencies and the phase of the normal-branch oscillations (NBOs) \citet{Yu+2001}

The time-evolution of the system  can be studied directly from equations (\ref{eq:govx}) and (\ref{eq:govg}) or with the aid of another integral of motion
\begin{eqnarray}
  F &=&
  \xi\Big[-\frac{8\sigma}{E}+\mu_\low\,\xi+\frac{\mu_\up}{\nu}(2-\xi)
  +\beta(1-\xi)(\xi E)^{1/2}\cos\gamma\Big].
\end{eqnarray}

Different solutions trace different curves $F(\gamma,\xi)=\mathrm{const}$ in the $(\gamma,\xi)$-disk. For given values of the parameters $\beta$, $\mu_\low$, $\mu_\up$, $E$ and $\sigma$ there may exist several fixed points in this disk, each representing strictly periodic oscillations of the system. They correspond to oscillations whose amplitudes are not modulated and whose phases $\phi_\low(t)$ and $\phi_\up(t)$ are linear functions of time that adjust the frequencies of oscillations to the exact 3:2 ratio. The fixed points are either extrema or saddles of the functions $F(\xi,\gamma)$ and their positions in the $(\gamma,\xi)$-disk are given by the conditions $\dot{\xi}=\dot{\gamma}=0$  imposed on equations (\ref{eq:govx}) and (\ref{eq:govg}). Generally, their $\gamma$-coordinate equals $k\pi$ with $k$ being an integer. 

\begin{figure*}
  \begin{center}
    \includegraphics[width=0.30\textwidth]{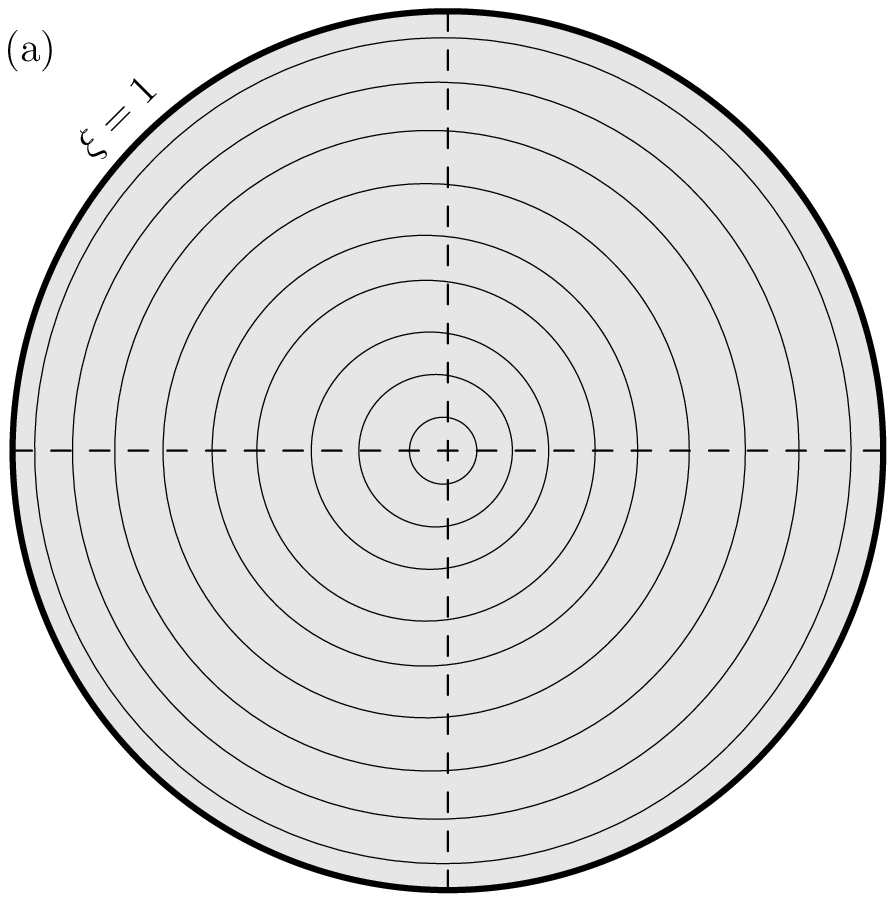}
    \hfill
    \includegraphics[width=0.30\textwidth]{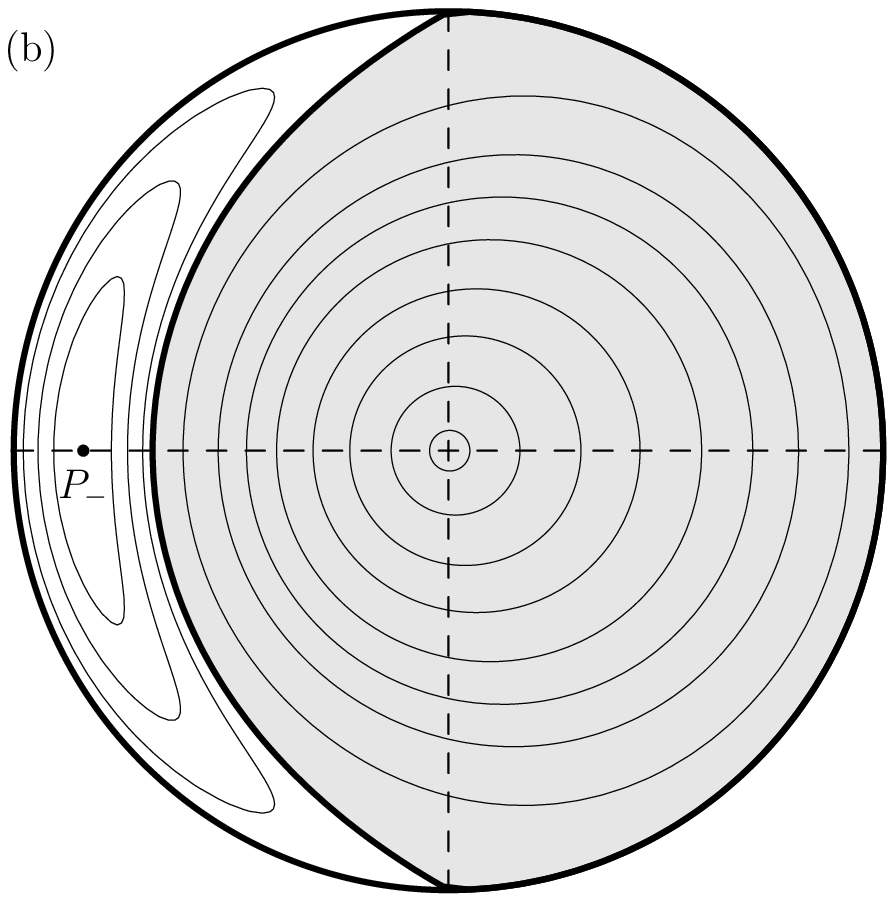}
    \hfill
    \includegraphics[width=0.30\textwidth]{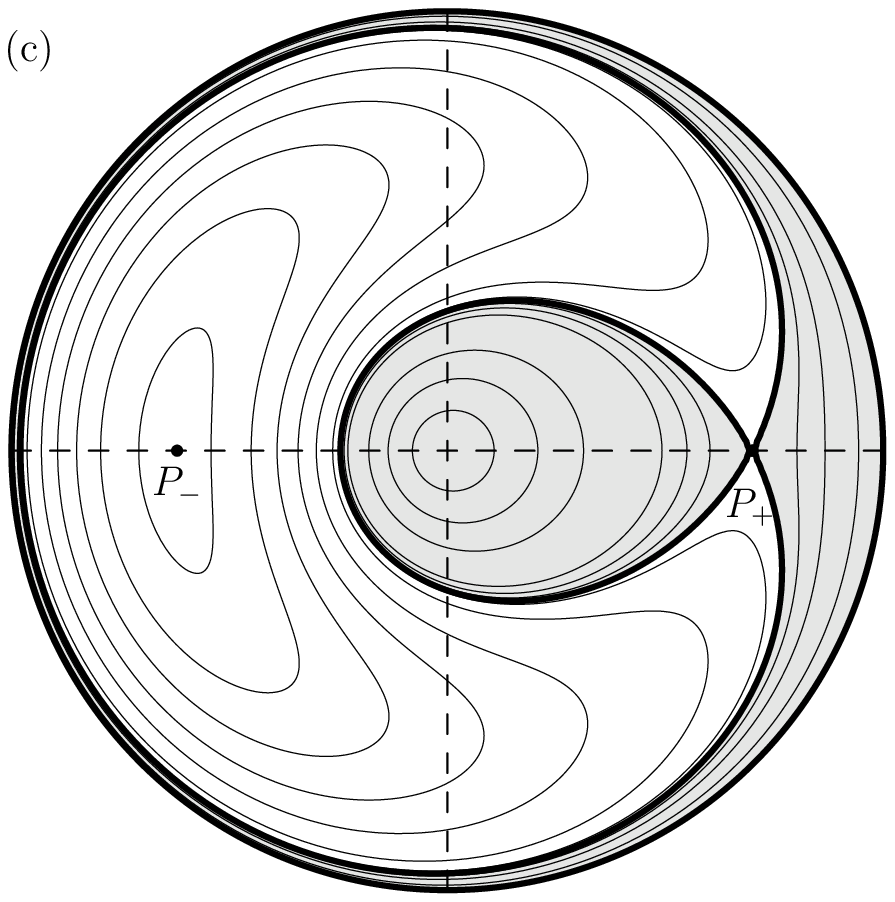}
  \end{center}
  \caption{Three possible types of topologies of the $(\gamma,\xi)$-disk denoted in Fig.~\ref{fig:tongue} as A, B and C (left, middle and right, respectively). The polar angle and the distance from the origin represent the phase function $\gamma$ and the fractional energy $\xi$, respectively. The values of the total energy and detuning parameter corresponding to these examples are $(\sigma/\omega_\up, E) = (0.005, 0.1)$, $(0.025, 0.1)$ and $(0.02, 0.1)$. Each curve corresponds to a solution of the governing equations (\ref{eq:govx}) and (\ref{eq:govg}). The disks corresponding to $(\sigma,E)$ from the resonance tongues contain both the librating (white regions), and circulating (shaded regions) trajectories. The fixed points $P_\pm$ correspond to a saddle point or extrema of the integral of motion $F(\xi,\gamma)$.}
  \label{fig:gxplanes}
\end{figure*}

The number of fixed points in the $(\gamma,\xi)$-disk depends on the energy $E$ and on the detuning parameter $\sigma$. For a fixed value of $\omega_\up$, the latter quantity is given by the ratio of the eigenfrequencies, $\mathcal{R}\equiv\omega_\up/\omega_\low$. Hence, it is possible to indicate regions in the $(\mathcal{R},E)$ plane for which different number of fixed points exists. Figure~\ref{fig:tongue} shows these regions for the system described by the fiducial set of parameters, $\beta=\mu_{\ell}=\mu_\up=1$, $\nu=9/4$. 

Some examples of the three possible topologies of the $(\gamma,\xi)$-disk are given in Figure~\ref{fig:gxplanes}.
All of them contain circulating trajectories for which  $\gamma(t)$ monotonically increases or decreases in the full range (\eg $0\leq\gamma<2\pi$). The ratio of observed frequencies can be expressed (with an accuracy of up to $\epsilon^3$) as 
\begin{equation}
  \mathcal{R}^\star\equiv\frac{\omega_\up^\star}{\omega_\low^\star}=
  \frac{3}{2}\left(1+\frac{\dot{\gamma}}{2\omega_\up}\right).
  \label{eq:R} 
\end{equation}
Hence, this ratio is always greater or smaller than 3/2  if the system evolves along the circulating trajectories with increasing or decreasing $\gamma$, respectively. 

In addition, closer to the resonance, the $(\gamma,\xi)$-disks contain a fixed point that corresponds to an extremum of the function $F(\gamma,\xi)$. This point is always surrounded by librating orbits with two turning points where  $\dot{\gamma}$ changes its sign. When the system follows these trajectories the observed frequency ratio oscillates around 3:2 . The regions of librating and circulating orbits are divided by a separatrix. When the $(\gamma,\xi)$-disk contains two fixed points, the second point is a saddle point, in which the trajectory crosses itself. Consequently, it takes infinite time for the system to reach the saddle point from any other point on the separatrix.

\section{Slow passage through the resonance}
Let us now suppose that the eigenfrequencies are tunable and are subject to small slow changes. More specifically, we consider the case when the eigenfrequencies and couplings are slowly changing and the system gradually passes through the resonance. We concentrate on that part of this process in which the eigenfrequencies are close to the rational 3:2 ratio. We assume that the total relative changes of the eigenfrequencies are small (of the order of $\epsilon^2$), and the timescale for this change, $T$, is much longer than any other timescale connected to the nonlinear oscillations (in particular $T\gg t_{~\mathrm{mod}}$).

In general, the oscillations of the system are the solutions to equations (\ref{eq:govl}) and (\ref{eq:govu}), where $\omega_{\low}$ and $\omega_\up$ as well as the nonlinear functions $f_\low$  and $f_\up$ are known functions that depend explicitly on time. Since the characteristic timescale of the process $T$ is much longer than $t_{~\mathrm{mod}}$, the only change in the equations (\ref{eq:govx})--(\ref{eq:govg}) is that the eigenfrequencies $\omega_\low$ and $\omega_\up$ and the coefficients $\beta$, $\mu_\low$, $\mu_\up$ and $\nu$ become slow functions of time. Close to the resonance, the oscillations are most sensitive to the change of the detuning parameter $\sigma(t)\equiv2\omega_\up(t)-3\omega_{\low}(t)$, while they are practically unaffected by small changes of the other quantities. This is because the coefficients of the other terms in equations (\ref{eq:govx}) and (\ref{eq:govg}) are always multiplied by small quantities (of the order of $\sim\epsilon^{~2}$ and $\sim\epsilon^{~3}$). We do not consider the ``singular" case when the two eigenfrequencies are nearly correlated as $\omega_\up\approx 1.5\omega_\low$. Because the characteristic timescale of passage through resonance is taken to be much longer than that of the resonant modulation we may neglect time derivatives of $\sigma$ and $E$ in the evolution equations (8) and (9) that describe the evolution of the system on the timescale $t_{~\mathrm{mod}}$.

As the system passes through the resonance, the topology of the $(\gamma,\xi)$-disk gradually changes. The system trajectory $[\gamma(t),\xi(t)]$ does not strictly follow the curves $F(\gamma,\xi)=\mathrm{const}$ because the ratio $\sigma/E\sim1$ is changing significantly. Indeed both quantities are of the order of $\epsilon^2$ and the change of the total energy $E$ is negligible because the relative change of $\nu$ is proportional to $\epsilon^2$. Accordingly, in the following, we take $\nu=\mathrm{const}$.

\begin{figure*}
  \begin{center}
    \hfill
    \includegraphics[width=0.40\textwidth]{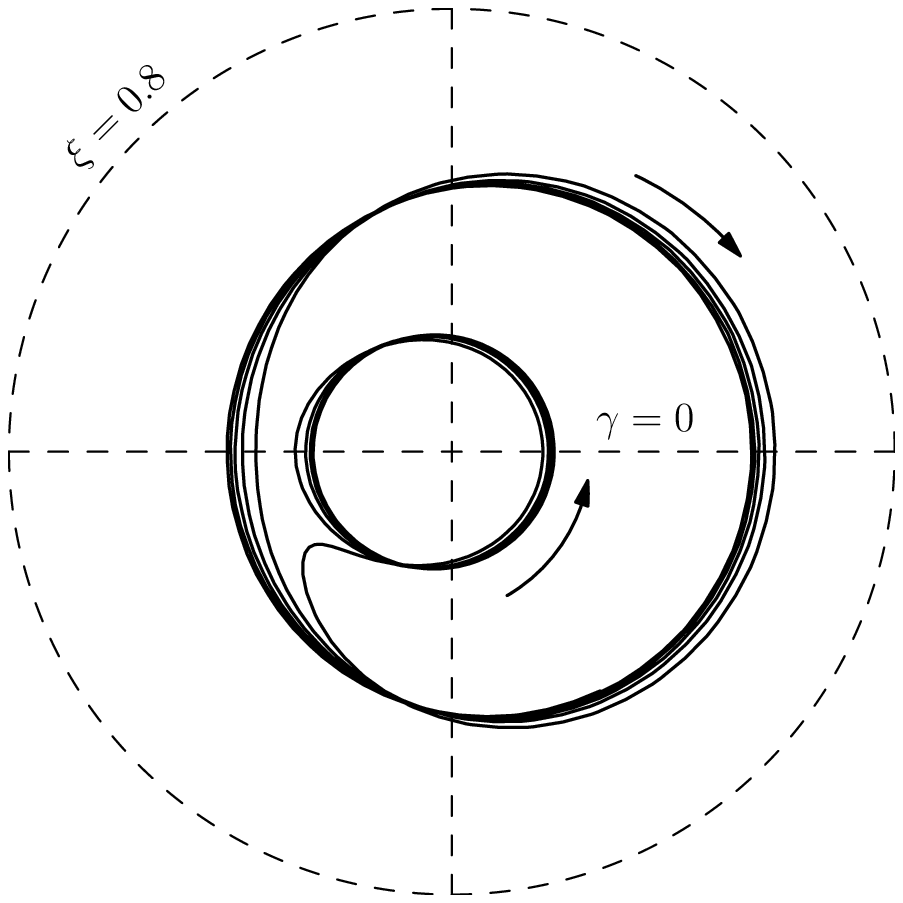}
    \hfill
    \includegraphics[width=0.50\textwidth]{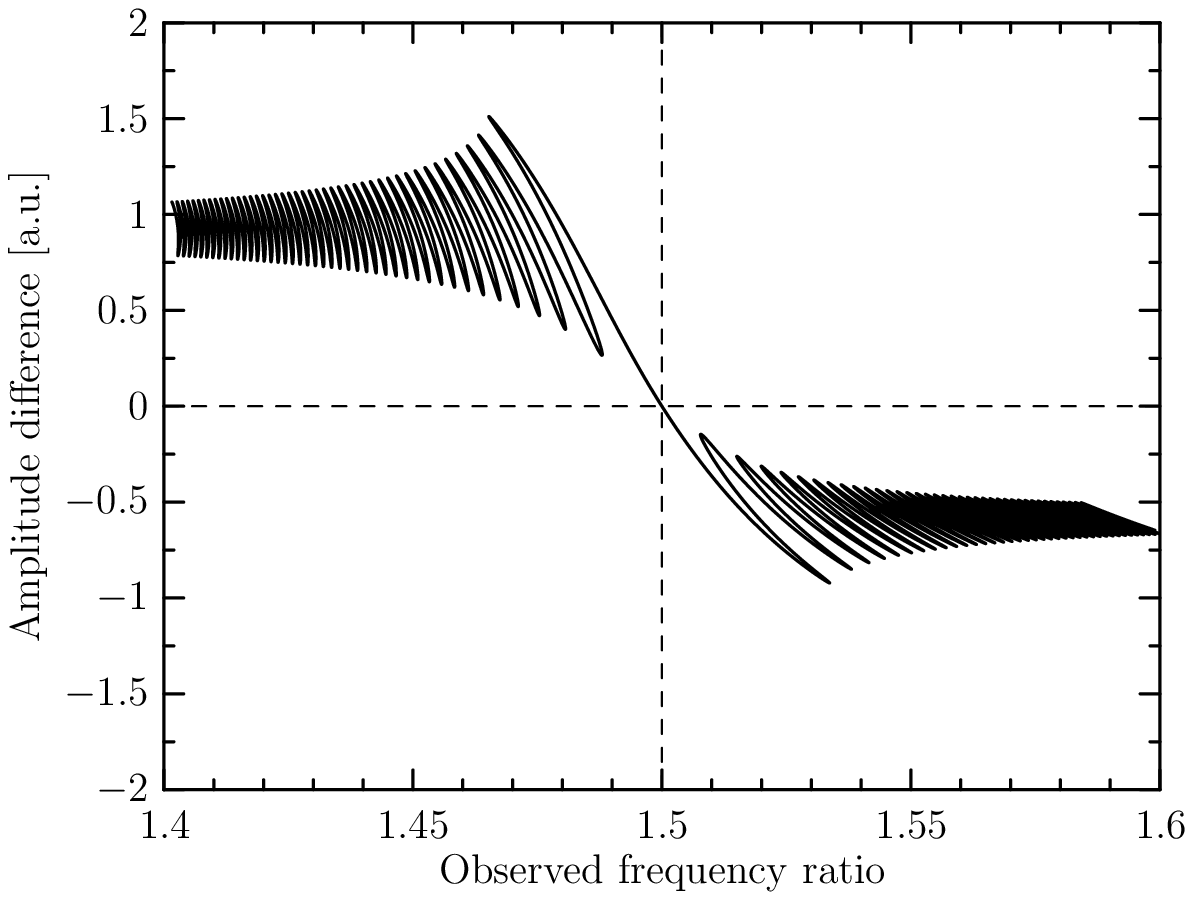}
  \end{center}
  \caption{Behavior of the resonant oscillations of the system undergoing 
  a slow secular change of its eigenfrequencies. The detuning parameter $\sigma$
  increases linearly with time so that the system gradually passes through  the resonance.
  The left panel shows the trajectory of the system projected into the $(\gamma,\xi)$-disk.
  The right panel shows behavior of the amplitude difference, $\delta a=a_\low-a_\up$,
  (measured in arbitrary units) as a function of the ratio the observed frequencies, 
  $\mathcal{R}^\star=\omega_\up^\star/\omega_\low^\star$. 
  Because of the slow changes in the eigenfrequencies, the energy
  exchange between modes is not exactly symmetric and the total energy is redistributed
  between the modes when the system passed the resonance.}
  \label{fig:transit}
\end{figure*}

The general effect of the resonance transition is illustrated by the example shown in Figure~\ref{fig:transit}. The eigenfrequencies are changed with constant rate, $\sigma(t) = \sigma_\mathrm{i} + \dot{\sigma}\,t$. The left panel shows the evolution of the oscillations in the $(\gamma,\xi)$-disk and the right panel shows the observable quantities (\ie the ratio of observed frequencies $\mathcal{R}^\star$ versus the amplitude difference $\delta a = a_\low-a_\up$). These Figures are for $\sigma_\mathrm{i} = -0.1\omega_\up$, $\dot{\sigma} = 3\times10^{-5}\omega_\up^2$,  where $\omega_\up$ is the upper frequency in the exact resonance. The system is described by $\beta=-\mu_\ell=\mu_\up=5$ and $\nu=9/4$, and the initial conditions are $\xi_\mathrm{i}\equiv\xi(0)=0.5$, $\gamma_\mathrm{i}\equiv\gamma(0)=0$. The total energy is fixed at $E=0.1$. We also tried different rates of passage and different initial conditions.

In all cases the behavior of the system is qualitatively similar: the initial value of the detuning parameter corresponds to a point $(\sigma_\mathrm{i}, E)$ outside the resonance tongue. Hence, the $(\gamma,\xi)$-disk initially contains only circulating orbits. One of them is closely followed by the system (the outer ``circles'' in the left panel in Figure 3). The $\xi$-coordinate slightly oscillates in correspondence to a weak energy exchange between the modes. As the system approaches the resonance, the magnitude of these oscillations increases gradually. Then the system enters into the resonance tongue and the topology of the $(\gamma,\xi)$-disk suddenly changes -- the librating  trajectories appear. When the trajectory of the system crosses the separatrix and enters the region of librating trajectories, the energy exchange becomes asymmetric -- i.e., the inflow of the energy into one of the modes exceeds the outflow. The system may escape from this region passing the separatrix again before it makes a single orbit (as in Fig.~\ref{fig:transit}) or it can be trapped in this region for longer time. A trapping in the region of librating orbits is accompanied by several changes of the sign of $\dot{\gamma}$, and therefore by changes of the direction of orbiting in the $(\gamma,\xi)$-disk. The ratio $\mathcal{R}^\star$ of the observed frequencies oscillates around 1.5 [as follows from equation (\ref{eq:R})]. When the system leaves the resonance tongues and as it gets farther away from them, the energy exchange between modes becomes more symmetric and less apparent. The final fractional energy $\xi_\mathrm{f}$ after the passage is generally different from the initial one $\xi_\mathrm{i}$: the energy is redistributed due to the resonance.

We note that the behavior of the amplitudes with respect to the frequency ratio as shown in the right panel of Figure~\ref{fig:transit} is similar to that reported for several atoll sources \citep{Torok2008}. 

\begin{figure*}
  \begin{center}
    \hfill
    \includegraphics[width=0.40\textwidth]{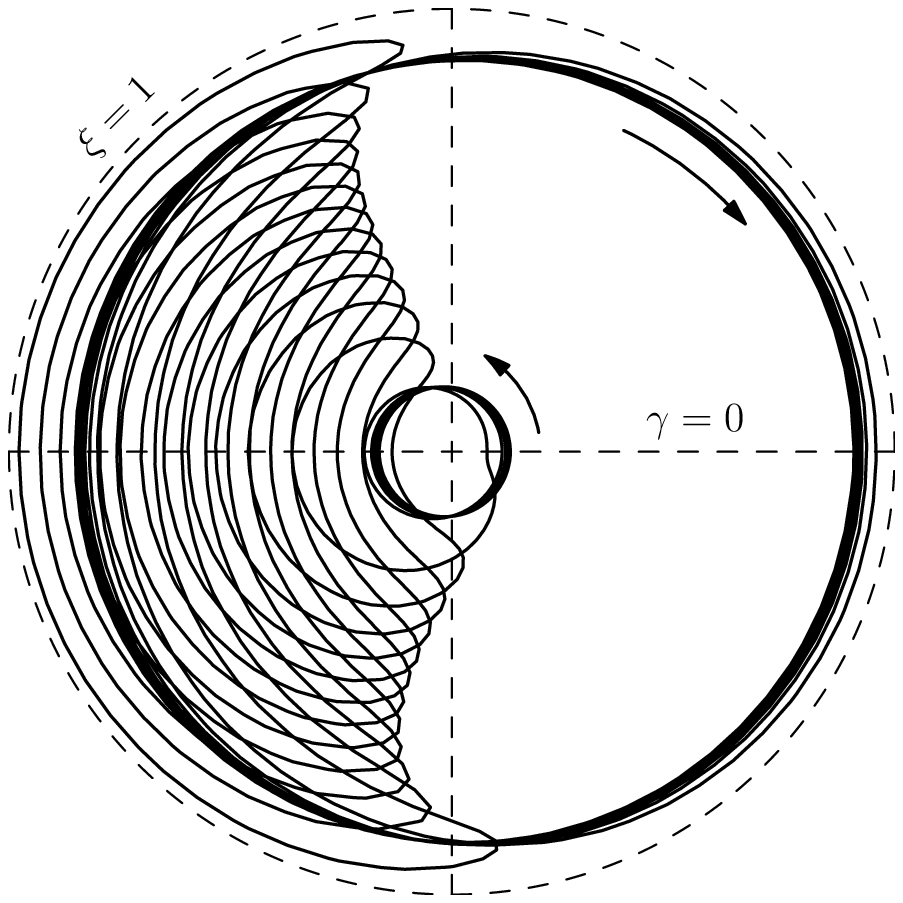}
    \hfill
    \includegraphics[width=0.50\textwidth]{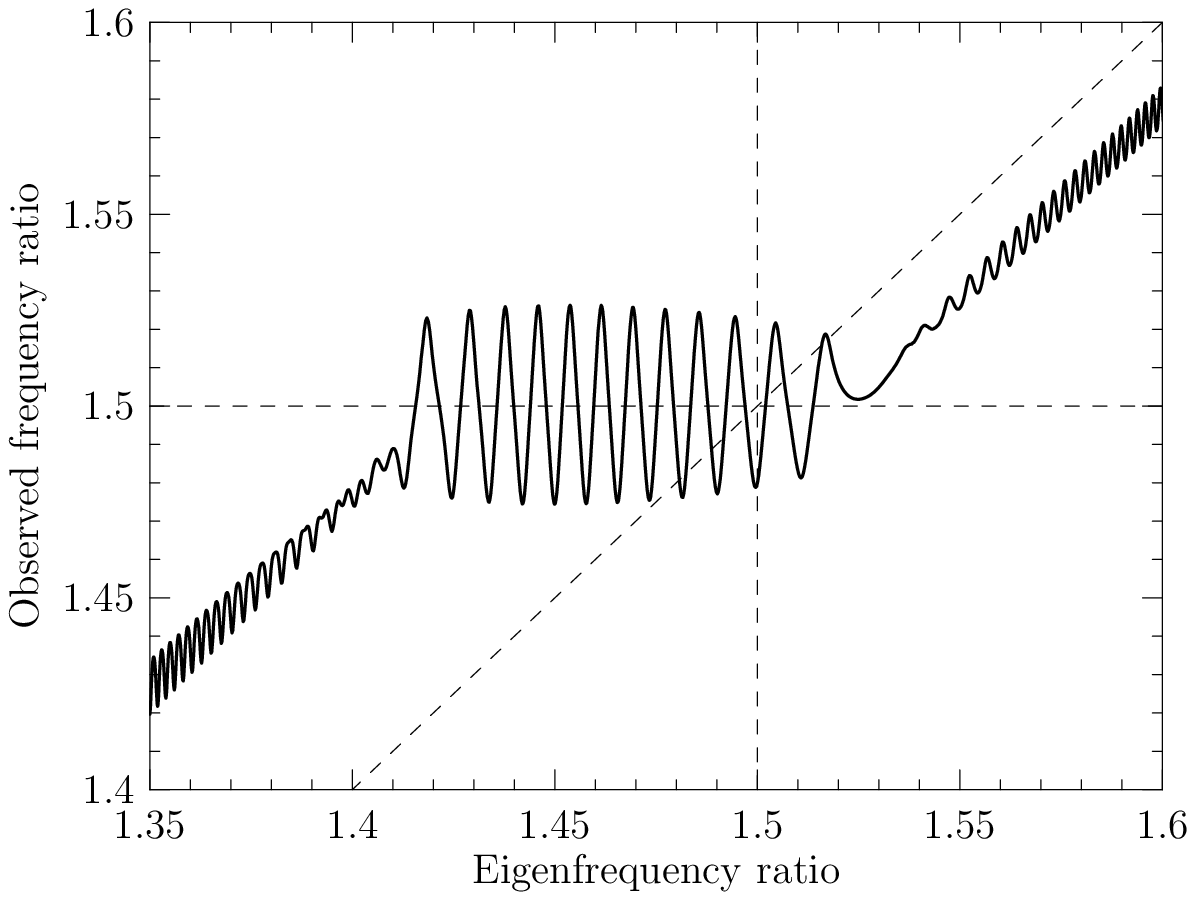}
  \end{center}
  \caption{Example of a trajectory with a frequency-locking episode. Left: Trajectory projected onto
  the $(\gamma,\xi)$-disk. frequency locking corresponds to the redistribution of the energy when
  the system is trapped in the region of the librating orbits. Right: The frequency-locking effect in the plot
  showing ratio of the observed frequencies versus eigenfrequency ratio.}
  \label{fig:locking}
\end{figure*}

An example of a trajectory trapped in the region of the librating orbits is shown in Figure~\ref{fig:locking}. The parameters of the solution are $\sigma_0 = -0.2\omega_\up$, $\dot{\sigma} = 4\times10^{-5}\omega_\up^2$, $\beta=\mu_\ell=-\mu_\up=5$ and $\nu=9/4$. The initial conditions are $\xi_\mathrm{i}=0.9$, $\gamma_\mathrm{i}=0$ and the total energy is $E=0.1$. Obviously, the nonlinearities adjust the observed frequencies so that their ratio oscillates around the resonance value 3:2 during the trapping period.

\begin{figure*}
 \begin{center}
   \includegraphics[width=0.49\textwidth]{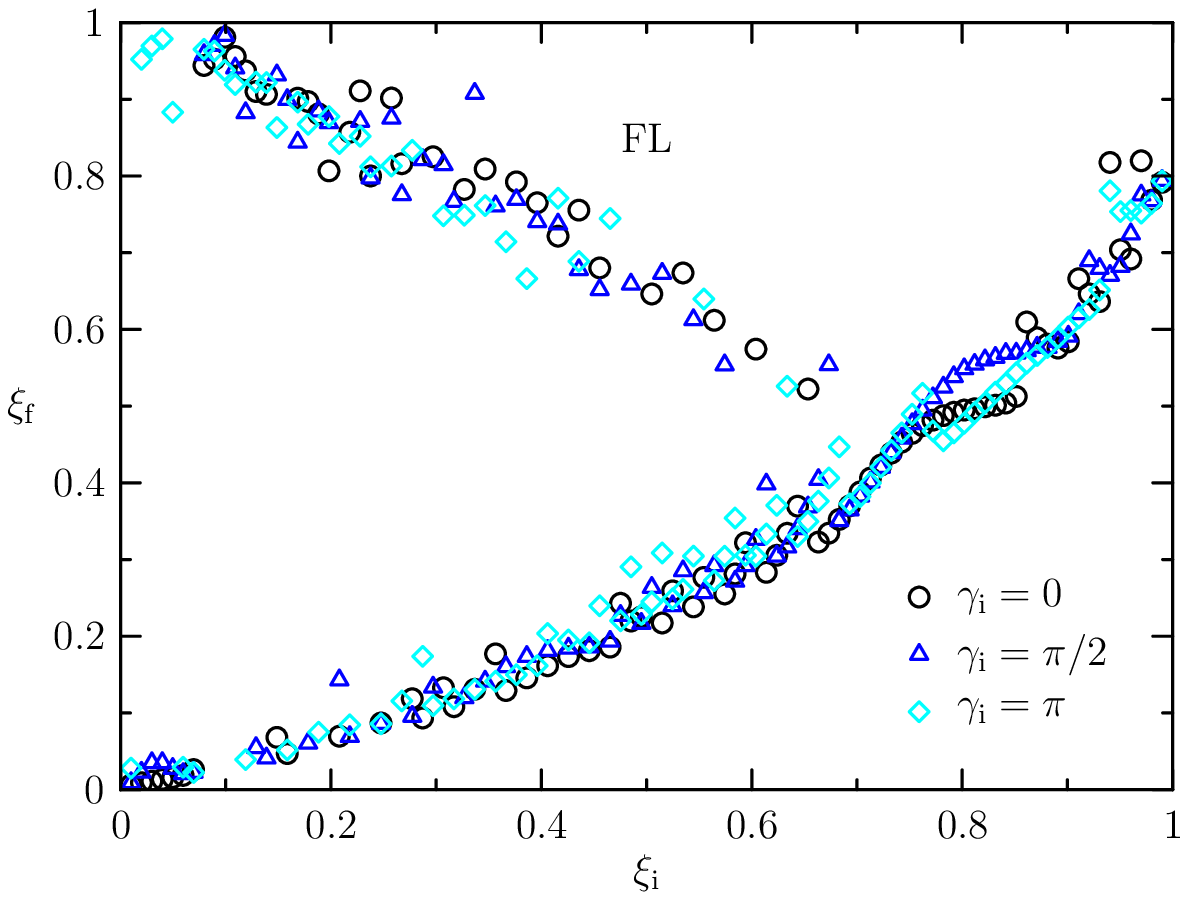}
   \hfill
   \includegraphics[width=0.49\textwidth]{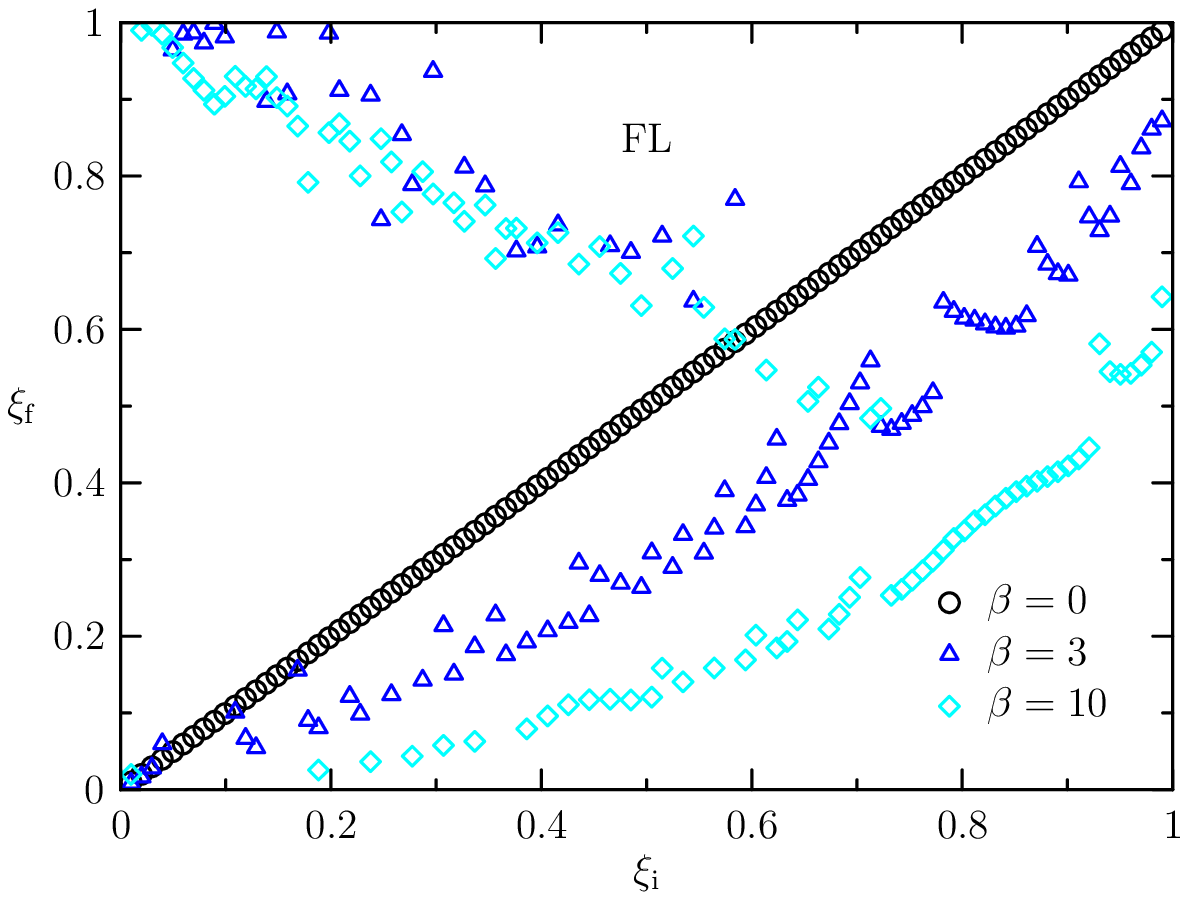}
 \end{center}
 \caption{Energy exchange during the forward transition through the 3:2 resonance 
 (\eg from the lower to the higher eigenfrequency ratios). $\xi_\mathrm{i}$ and 
 $\xi_\mathrm{f}$ are the fractional energy in the lower oscillation mode 
 before and after the transition. Left: Dependence on the initial condition (different symbols denote different initial phases, $\gamma_\mathrm{i}=0,\pi/2$ and $\pi$) for $\beta=5$. Right: Dependence on the value of the parameter $\beta$ (the initial phase is $\gamma_\mathrm{i}=0$). The branches corresponding to solutions with (denoted by `FL') and without the frequency-locking period are clearly separated. In this plot, $\xi_\mathrm{i}$ is sampled uniformly---an indication of the probability of finding the system in a `frequency-locked' solution is given by the relative number of points on the FL branch.}
 \label{fig:xi0f}
\end{figure*}

We examined the stability of this process against  changes of the initial conditions and other parameters characterizing the passage through the resonance. The effects of changing the initial conditions are summarized in the left panel of Figure~\ref{fig:xi0f}. Again, the parameters of the system were set to the values $\beta=-\mu_{~\ell}=\mu_\up=5$ and $\nu=9/4$ and the initial detuning parameter and the rate of change were $\sigma_{~0} = -0.1\omega_\up$, $\dot{\sigma} = 3\times10^{-5}\omega_\up$. The solutions trapped in the librating regions and showing for some time an episode of what could be termed `frequency-locking' are clearly separated from the others. These solutions are quite common for small $\xi_{~\mathrm{i}}$. In that case they are accompanied by significant exchanged energy. As the initial fractional energy $\xi_{~\mathrm{i}}$ increases they become rather exceptional. 

Although different for the solutions with and without frequency-locking, the $\xi_\mathrm{i}$--$\xi_\mathrm{f}$ dependence is fairly independent of the initial phase function $\gamma_\mathrm{i}$ as it can be seen from the left panel of Figure~5, where the solutions with different $\gamma_\mathrm{i}$ correspond to different symbols. Similarly, we also checked that the rate of the transition $\dot{\sigma}$ has no significant influence on the resulting $\xi_{~\mathrm{i}}$--$\xi_{~\mathrm{f}}$ plot.

The right panel of Figure~\ref{fig:xi0f} shows the effect of changing the value of the parameter $\beta$. The redistribution of energy in the solutions without the `frequency-locking' period becomes more pronounced as the  value of  $\beta$ increases. In addition, the solution with the `frequency locking' period become more frequent (see Table~\ref{tab:fl}). The value $\beta=0$ corresponds to the case when there is no resonant coupling between the two oscillations, and $\xi_\mathrm{i}=\xi_\mathrm{f}$. Of course, there are no `frequency-locked' solutions in this case.

The dependence on the remaining parameters could be discussed in a similar way; however, it can be shown that for sufficiently slow transitions of the resonance the whole dynamics depends only on a single parameter, $\hat{\beta} = \beta\sqrt{E}/(\mu_\low - \mu_\up/\nu)$. Therefore one does not expect any qualitatively new behavior as the other parameters are changed.

\begin{table}
  \caption{Occurrence of solution with frequency-locking period 
  in sets of 200 solutions for different values of 
  $\gamma_\mathrm{i}$ and $\beta$. In each set the initial 
  value $\xi_\mathrm{i}$ is uniformly distributed between 0 and 1, 
  while the other parameters are kept constant.}
  \label{tab:fl}
  \centering          
  \begin{tabular}{cccccc}   
    \hline\hline           
     & $\beta=0$ & $\beta=3$ & $\beta=5$ & $\beta=10$ \\ 
    \hline
    $\gamma_\mathrm{i}=0$ & 0 & 44 (22\%) & 64 (32\%) & 89 (48\%) \\
    $\gamma_\mathrm{i}=\pi/2$  & 0 & 47 (24\%) & 60 (30\%) & 95 (48\%) \\
    $\gamma_\mathrm{i}=\pi$    & 0 & 43 (22\%) & 60 (30\%) & 88 (44\%) \\
    \hline\hline                
  \end{tabular}
\end{table}

\section{Discussion and conclusions}
\label{sec:concl}
If the  twin kHz QPOs in neutron-stars correspond to two physical oscillators that have a nonlinear coupling, the two QPOs may exhibit resonance phenomena when the QPO frequencies are close to a ratio of small integer numbers. Among the most characteristic observable imprints of the internal resonance are (i) Exchange of energy between the oscillations (ii) equality of the quality factors of oscillations and (iii) locking of the frequency ratio of oscillations. Our work was devoted to the mathematics of the first and third phenomena. 

Some models of QPOs, which may support an internal resonance of the type described in this work have been described in Section 1. In this work we explored only the 3:2 resonance which is the one most prominent in the data \citep{Abramowicz+2003a, Torok+2008}. The examination of other resonances may be done in an analogous way.

Assuming a slow (secular) change of the eigenfrequencies of the two coupled oscillators, we demonstrated that the energy is redistributed between the two oscillators when the system passes through a resonance. This effect can be seen in the change of the sign in the amplitude difference when the ratio of the eigenfrequencies ratio crosses a rational value, such as 3:2. This result is in qualitative agreement with the observations for some neutron stars \citep{Torok2008}.

A class of solutions exhibits a time interval during which  a frequency-locking of the two oscillations occurs. This corresponds to small oscillations of the actual frequencies around the ratio 3:2 over a (greater) range of eigenfrequency ratios (Figure~4). 

\acknowledgements
This work was started at the G\"oteborg University and continued at MPA Garching. JH  is greatful to prof. Chengmin Zhang for useful discussions and acknowledges support of the grants GA\v{C}R 205/07/0052 and GA\v{C}R 205/06/P415, PR is supported by the Pappalardo Fellowship at MIT and MAA was supported by the Polish Ministry of Education grant N203 009 31/1466, WK acknowledges partial support through grant 1P03D00530 and GT was supported by the grant MSM4781305903.


\bibliographystyle{aa}
\bibliography{cross}


\end{document}